\documentclass[12pt]{article}
\usepackage{amsmath,XoohmE}
\usepackage{graphicx,color}

\def\rA{{\rm A}}
\def\ah{{\hat a}}
\def\bh{{\hat b}}
\def\rB{{\rm B}}
\def\cB{\mathcal{B}}
\def\sC{\mathscr{C}}
\def\Dh{{\hat\Delta}}

\def\sF{\mathscr{F}}
\def\rI{{\rm I}}
\def\hi{{\hat\imath}}
\def\iW{\i_{\sss W}}
\def\rJ{{\rm J}}
\def\bJ{\boldsymbol{\J}}
\def\hj{{\hat\jmath}}
\def\hk{{\hat k}}

\def\rL{\text{L}}
\def\sL{\mathscr{L}}

\def\hl{{\hat\ell}}

\def\sign{\mathop{\text{sign}}\nolimits}

\def\sV{\mathscr{V}}
\def\cW{\mathcal{W}}

\def\bX{\boldsymbol{X}}

\def\ddt{\partial_\t}

\def\bsm{\left[\begin{smallmatrix}}
\def\esm{\end{smallmatrix}\right]}

\def\vC#1{\vcenter{\hbox{\hss#1\hss}}}
\def\InT#1{\\[-4mm]\intertext{#1\vspace{-4mm}}}

\definecolor{Hey}{rgb}{1,0,.4}
\definecolor{Redd}{rgb}{.7,.05,.1}
\definecolor{orange}{rgb}{.8,.4,0}
\definecolor{plum}{rgb}{.4,0,.6}
\definecolor{gold}{rgb}{.9,.75,0}
\definecolor{beig}{rgb}{1,.9,.75}
\definecolor{bloo}{rgb}{0,.15,.25}
\definecolor{gray}{rgb}{.6,.6,.6}

\catcode`@=11
\def\Ft#1{\,\footnote{#1}}
\newdimen\parshift\parshift=\parindent
 \long\def\@footnotetext#1{\insert\footins{\reset@font\footnotesize
           \interlinepenalty\interfootnotelinepenalty\splittopskip%
            \footnotesep\splitmaxdepth\dp\strutbox\floatingpenalty\@MM%
             \hsize\columnwidth\addtolength{\hsize}{-2\parindent}
              \@parboxrestore\protected@edef\@currentlabel%
              {\csname p@footnote\endcsname\@thefnmark}%
                \color@begingroup%
                 \@makefntext{\rule\z@\footnotesep\ignorespaces#1%
                  \@finalstrut\strutbox}%
                \color@endgroup}}
 \long\def\@makefntext#1{\hglue\parshift%
           \vbox{\noindent\hb@xt@0em{\hss\@makefnmark\kern1pt}#1}}
\catcode`@=12
 \font\rOpe=cmsy10                        
 \def\ktl{{\hbox{\rOpe\char'170}}}        
 \def\kbl{{\hbox{\rOpe\char'170}}}        
 \def\kcr{{\reflectbox{\rOpe\char'170}}}  
 \def\ktr{{\reflectbox{\rOpe\char'170}}}  
 \def\kbr{{\reflectbox{\rOpe\char'170}}}  
 \def\Border{\vbox{\hsize0pt
        \setlength{\unitlength}{1mm}
        \newcount\xco
        \newcount\yco
        \xco=-21
        \yco=12
        \begin{picture}(0,0)(-7.5,0)
        \put(\xco,\yco){$\ktl$}
        \advance\yco by-1
        {\loop
        \put(\xco,\yco){$\kcr$}
        \advance\yco by-2
        \ifnum\yco>-240
        \repeat
        \put(\xco,\yco){$\kbl$}}
        \xco=170
        \yco=12
        \put(\xco,\yco){$\ktr$}
        \advance\yco by-1
        {\loop
        \put(\xco,\yco){$\kcr$}
        \advance\yco by-2
        \ifnum\yco>-240
        \repeat
        \put(\xco,\yco){$\kbr$}}
        \put(-19.5,13){\scalebox{.598}{State University of New York
            Physics Department|University of Maryland Center for
            String and Particle  Theory \&\ Physics Department|%
            Delaware State University DAMTP}}
        \put(-19.5,-241.5){\scalebox{.728}{University of Washington
            Mathematics Department|Pepperdine University Natural
            Sciences Division|Bard College Mathematics
            Department}}
        \end{picture}
        \par\vskip-8mm}}
\definecolor{UMred}{rgb}{.9,.05,.2}
 \def\UMbanner{\vbox{\hsize0pt
        \setlength{\unitlength}{.4mm}
        \thicklines
        \begin{picture}(0,0)(-30,-10)
        \put(165,16){\line(1,0){4}}
        \put(170,16){\line(1,0){4}}
        \put(180,16){\line(1,0){4}}
        \put(175,0){\line(1,0){4}}
        \put(180,0){\line(1,0){4}}
        \put(185,0){\line(1,0){4}}
        \put(169,0){\line(0,1){16}}
        \put(170,0){\line(0,1){16}}
        \put(179,0){\line(0,1){16}}
        \put(180,0){\line(0,1){16}}
        \put(184,0){\line(0,1){16}}
        \put(185,0){\line(0,1){16}}
        \put(169,16){\oval(8,32)[bl]}
        \put(170,16){\oval(8,32)[br]}
        \put(179,0){\oval(8,32)[tl]}
        \put(185,0){\oval(8,32)[tr]}
        \end{picture}
        \par\vskip-6.5mm
        \thicklines}}

 \def\brr{\begin{eqnarray}}
 \def\err{\end{eqnarray}}

 \SfTitles 
 \allowdisplaybreaks
\begin{document}

\thispagestyle{empty}
\vbox{\Border\UMbanner}
 \noindent
 \today\hfill{UMDEPP 07-015}
  \vfill
\setcounter{page}{0}\thispagestyle{empty}
 \begin{center}
{\LARGE\sf\bfseries\boldmath
Super-Zeeman Embedding Models\\[3mm]
on $N$-Supersymmetric World-Lines}\\[10mm]
  \vfill
{\sf\bfseries C.F.\,Doran$^a$, M.G.\,Faux$^b$, S.J.\,Gates, Jr.$^c$,
     T.\,H\"{u}bsch$^d$, K.M.\,Iga$^e$ {\rm and} G.D.\,Landweber$^f$}\\[2mm]
{\small\it
  $^a$Department of Mathematics,\\[-1mm]
      University of Washington, Seattle, WA 98105%
  \\[-4pt] {\tt  doran@math.washington.edu}
  \\
  $^b$Department of Physics,\\[-1mm]
      State University of New York, Oneonta, NY 13825%
  \\[-4pt] {\tt  fauxmg@oneonta.edu}
  \\
  $^c$Center for String and Particle Theory,\\[-1mm]
      Department of Physics, University of Maryland, College Park, MD 20472%
  \\[-4pt] {\tt  gatess@wam.umd.edu}
  \\
  $^d$Department of Physics \&\ Astronomy,\\[-1mm]
      Howard University, Washington, DC 20059, and\\[-1mm]
  Department of Applied Mathematics and Theoretical Physics,\\[-1mm]
      Delaware State University, Dover, DE 19901
  \\[-4pt] {\tt  thubsch@desu.edu}
  \\
  $^e$Natural Science Division,\\[-1mm]
      Pepperdine University, Malibu, CA 90263%
  \\[-4pt] {\tt  Kevin.Iga@pepperdine.edu}
  \\
 $^f$Department of Mathematics, Bard College,\\[-1mm]
     Annandale-on-Hudson, NY 12504-5000%
  \\[-4pt] {\tt  gregland@bard.edu}
 }\\[5mm]
  \vfill\vfill
{\sf\bfseries ABSTRACT}\\[3mm]
\parbox{5.1in}{
We construct a model of an electrically charged magnetic dipole with arbitrary $N$-extended world-line supersymmetry, which exhibits a supersymmetric Zeeman effect.
 By including supersymmetric constraint terms, the ambient space of the dipole may be tailored into an algebraic variety, and the supersymmetry broken for almost all parameter values. The so exhibited obstruction to supersymmetry breaking refines the standard one, based on the Witten index alone.}

\end{center}
 \vfill\vfill\vfill

\clearpage
\section{The 1D, $N=1$ Super-Zeeman Embedding Model}
Quantum mechanics with $N$-extended supersymmetry has been a topic of our recurrent interest\cite{rGR0,rGR1,rGR2,rGLPR,rGLP,rA,r6-1}. Here, we construct non-trivial quantum-mechanical models invariant with respect to {\em\/arbitrarily high\/} $N$-extended supersymmetry, generated by $N$ supercharges, $Q_\rI$:
\begin{alignat}{3}
 \big\{\,Q_\rI\,,\,Q_\rJ\,\big\} &= 2i\,\d_{\rI\rJ}\,\ddt~, \quad&\quad
 \big[\, Q_\rI\,,\,\ddt\,\big] &= 0~,\qquad
 \rI,\rJ=1,\cdots\,N\in\IN~. \label{eSuSy}
\end{alignat}
In particular, this class of models exhibits: ({\bf1})~coupling to external magnetic fields, ({\bf2})~target space embedding as algebraic varieties, and ({\bf3})~supersymmetry breaking by constraint geometry.

To motivate our construction and its generalizations, we first discuss its $N=1$-supersymmetric toy model version. This model includes {\em\/several\/}, and by no means all the features with which one may wish to endow it. As our present purpose is the generalization to $N>1$, we focus on the select features highlighted above, and defer both other generalizations and most specializations to particular subregions in the multi-dimensional parameter space to a subsequent effort.

\subsection{Angular Momentum and External Magnetic Fields}
A charged particle moving in a two-dimensional, $(x,y)$-plane may possess:
\begin{alignat}{3}
 &\text{angular momentum}:&\qquad
 \rL &:=\, m ( x\,\dot y\,-\,y\,\dot x )~,\qquad\qquad \label{eL}\\
 &\text{magnetic dipole moment}:&\qquad
 \mu     &:=\,\frac{q_0}{2 m c}\,\rL~. \label{emu}
\end{alignat}
An external, constant magnetic field $\cB_0$ that couples to this magnetic dipole moment contributes to the total energy of this particle through the well-known dipole term, equal to $\mu\cB_0\cos\q$, where $\q$ is the angle between $\cB_0$ and the normal to the $(x,y)$-plane.

This raises the obvious question: ``Is there a generalization of this interaction with an external magnetic field, which is invariant with respect to {\em arbitrarily $N$-extended\/} supersymmetry?''

To answer this question, we start with manifest $N=1$ supersymmetry, and introduce two real superfields, $\bX^a$ with $a=1,2$, with component fields:
\begin{equation}
 x^a:=\bX^a|~,\qquad\text{and}\qquad
 \c^a\Defl iD\bX^a|~,\qquad
 \text{with}\quad x^1=x~,~~x^2=y~, \label{eCmpX}
\end{equation}
where the trailing ``$|$'' denotes the evaluation of the preceding superfield expression by setting the Grassmann coordinates to zero , and the factor of $i$ ensures that $\c^a$ too are real. The supersymmetry transformation rules may be written as
\begin{equation}
 Q\, x^a = \c^a~,\qquad\text{and}\qquad
 Q\,\c^a = i\,\dot x^a~.\label{eISN1}
\end{equation}

With these, we note that the angular momentum, L, is part of the ``top'' component of a superfield expression:
\begin{align}
 \ve_{ab}\,D\,\big(\,\bX^a\, D\bX^b\,\big)\big|
   ~=~ -i\,\Big(\,\frac\rL{m}  ~-~ 2i\,\c^1\,\c^2 \,\Big)~.
\label{eN1wIb}
\end{align}
 As supersymmetry transforms the ``top'' component of any superfield expression into a total $\t$-derivative,
\begin{equation}
\sL_\text{L}
 ~=~ i\,\Big(\frac{q_0\,\cB_0}{2\,c}\cos\q\Big)\,
  \ve_{ab}\,D \big(\, \bX^a D\bX^b\,\big) \big|
 ~=~\mu\,\cB_0\cos\q -  2i\,\Big(\frac{q_0\,\cB_0}{2\,c}\cos\q\Big)\,\c^1\,\c^2
\label{eN1uIb}
\end{equation}
is the $N=1$ supersymmetrization of the dipole-interaction term, $\m\,\cB_0\cos\q$.
 Before proceeding, we set $c,q_0\to1$, so that the Larmor frequency becomes $\w_L:=\frac{q_0\cB_0}{mc}\cos\q\to(\cB_0\cos\q)/m$. We also rescale all fields by $\sqrt{m}$, so that $m$ disappears from the Lagrangian. This fixes the engineering dimensions:
\begin{equation}
 [\,x^a\,]=[\,\bX^a\,]=-\inv2~,\qquad[\,\c^a\,]=0~,
 \qquad\text{and}\qquad[\,\w_L\,]=1~, \label{eBcD}
\end{equation}
and turns\eq{eN1uIb} into:
\begin{align}
 \sL_\text{L}
  &= \inv2\,\w_L\,\ve_{ab}\big[\, x^a\dot{x}^b - i\,\c^a\,\c^b\,\big]~
   = \inv2\,\w_L\,\big[\, (x\dot{y}-y\dot{x}) - 2\,i\,\c^1\,\c^2\,\big]~.
  \label{eLdip}
\end{align}
 The factor of two which appears multiplying $\c^1\,\c^2$ in the final term in\eq{eLdip} may be identified with the Land\'e $g$-factor, $g_s=2$, for spin-$\inv2$ particles.

We will also need fermionic superfields $\bJ^\rA=(\j^\rA\mid F^\rA)$, $\rA=0,1,2$, with components
\begin{equation}
 \j^\rA\Defl\bJ^\rA \big|~, \qquad  F^\rA\Defl D\bJ^\rA\big|~,\label{eCmpJ}
\end{equation}
the supersymmetry transformations of which may be written as:
\begin{equation}
 Q\,\j^\rA = i\,F^\rA~,\qquad\text{and}\qquad
 Q\,F^\rA  = \dot \j^\rA~, \label{eN1F'}
\end{equation}
Here, $\j^\rA$ denote fermions, and $F^\rA$ are bosons, and
\begin{equation}
 [\bJ^\rA]=[\j^\rA]=0~, \qquad\text{and}\qquad [F^\rA]=+\inv2~. \label{eFcD}
\end{equation}

\subsection{The Toy Model Lagrangian}
The toy model Lagrangian for this spinning, charged particle, with $a,b=1,2$, is:
\begin{subequations}
 \label{eSZEM1}
\begin{alignat}{3}
 \sL_\text{SZEM}&=\sL_\text{B} +  \sL_\text{L} +  \sL_\text{F}
                 + \sL_\text{B$\cdot$F} + \sL_\text{C}~,
  \label{eSZEMp}
\intertext{where the summands are as follows:}
 \sL_\text{B}
  &= -\inv2\,\d_{ab}\,D\,\big[\,(D\bX^a)(D^2\bX^b)\,\big]\big|
  ~=~\inv2\,\d_{ab}\,(\,\dot{x}^a\,\dot{x}^b+i\c^a\dot\c^b\,)~,
 \label{eLB} \\[2mm]
 \sL_\text{L}
  &= -\inv2\,\w_L\,\ve_{ab}\,D\,\big[\,\bX^a\,(D\bX^b)\,\big]\big|
  ~=~\inv2\,\w_L\,\ve_{ab}(\,x^a\,\dot{x}^b - i\c^a\c^b\,)~,
 \label{eLL}
\intertext{provide the standard kinetic terms for the $(x^a|\c^a)$ supermultiplets, and their $\w_L$-dependent bilinear interaction term, respectively.}
 \sL_\text{F}~
  &=~\inv2\,\ha\d_{\rA\rB}\,D\,\big[ \bJ^\rA(D\,\bJ^\rB)\big ]\big|
   ~=~\inv2\,\d_0\,(\,F^0 F^0+i\j^0\,\dot\j^0\,)
      +\inv2\,\d_{ab}\,(\,F^a F^b+i\j^a\,\dot\j^b\,)~,
  \label{eLF} 
\intertext{provides the standard kinetic terms for the $(\j^\rA|F^\rA)$ supermultiplets, and}
 \sL_\text{B$\cdot$F}&= \w_0\,D~\big[\, \d_{ab}\>\bJ^a\bX^b \,\big]\big|
  =\w_0\,\d_{ab}(\,F^a\,x^b~+~i\,\j^a\,\c^b\,)~,
 \label{eLBF}
\intertext{provides the $\w_0$-dependent mixing between $\bX^a$ and $\bJ^a$. Note that $\bJ^0$ is omitted from $\sL_\text{B$\cdot$F}$, but turns up in}
\sL_\text{C}
 &=\inv2\,g_0\,D~ \big[\,\bJ^0(\,\bX^a\,h_{ab}\,\bX^b-R^2\,)\big]\big|
 =g_0\,\big[\,\inv2\,F^0(x^a\,h_{ab}\,x^b-R^2)
                       +i\,\j^0\,(x^a\,h_{ab}\,\c^b)\,\big]~.
\label{eLC}
\end{alignat}
\end{subequations}

\subsection{The Toy Model Parameter Space and Features}
The real parameters occurring in the Lagrangian\eq{eSZEM1} have the following engineering dimensions:
\begin{equation}
 [g_0]=\frc32~,\qquad
 [\w_L]=[\w_0]=1~,\qquad
 [\d_0]=[h_{ab}]=0~,\qquad
 [R]=-\inv2~.
 \label{ePcD}
\end{equation}
The parameters $\w_L$, $\w_0$ and $g_0$ may be used selectively, to turn on/off the Lagrangian terms\eq{eLL}, \eq{eLBF} and\eq{eLC}, and $\d_0\to0$ in\eq{eLF} turns $(\j^0|F^0)$ into Lagrange multipliers.

\subsubsection{Decoupling Limit}
When $\w_0,g_0\to0$, the $\bX^a$ decouple from the $\bJ^\rA$. Their dynamics is governed by $\sL_\text{B}+\sL_\text{L}$, as given in \Eqs{eLB}{eLL}, and $\sL_\text{F}$ given in \Eq{eLF}, respectively.

\paragraph{The $\bX^a$:}
The Lagrangian\eqs{eLB}{eLL} produces the coupled equations of motion:
\begin{alignat}{3}
 \ddot x\,-\,\w_L\,\dot y&=0~, &\mkern120mu \dot\c^1\,+\,\w_L\,\c^2&=0~, \label{eZIn1}\\*
 \ddot y\,+\,\w_L\,\dot x&=0~, &            \dot\c^2\,-\,\w_L\,\c^1&=0~. \label{eZIn2}
\end{alignat}
This result is fairly standard for massive fermions, but not so for bosons: To see this, we ``diagonalize'' \Eqs{eZIn1}{eZIn2}:
\begin{alignat}{3}
 [\,\ddt^2+\w_L^2\,]\,\ddt\, x  &=0~,&\mkern120mu
 [\,\ddt^2+\w_L^2\,]\,\c^1
&=0~, \label{efj}\\*
 [\,\ddt^2+\w_L^2\,]\,\ddt\, y &=0~, &\mkern120mu
 [\,\ddt^2+\w_L^2\,]\,\c^2 &=0~.
\label{efc}
\end{alignat}
Indeed, the so-obtained Klein-Gordon equation for $\c,\h$ is the standard result, but the corresponding 3rd order differential equations for the bosons are not. Nevertheless, no unwelcome higher-derivative effect ensues: For any $\w_L \neq0$, the solutions of\eqs{eZIn1}{eZIn2} are
\begin{align}
  x(\t) &= x_+\cos(\w_L\t)+ x_-\sin(\w_L\t)+ x_0~,&
  \c^1(\t) &=\c_+\cos(\w_L\t)+\c_-\sin(\w_L\t)~, \label{eSfj}\\
   y(\t) &=-x_-\cos(\w_L\t)+ x_+\sin(\w_L\t)+ y_0~,&
  \c^2(\t) &=\c_-\cos(\w_L\t)-\c_+\sin(\w_L\t)~. \label{eSfc}
\end{align}
 This leaves four bosonic, $( x_+, x_-, x_0, y_0)$, and two fermionic, $(\c_+,\c_-)$, integration constants to be determined by initial and/or boundary conditions. Now, only $ x_0$ and $ y_0$ parametrize zero-modes, the remaining, bosonic-fermionic constant pairs are associated with the $\w_L$-modes. Thus, the Witten index is\Ft{Rotational symmetry implies that an additional bosonic and an additional fermionic mode within\eqs{eSfj}{eSfc} have zero energy. However, they leave the Witten index unchanged, and unless $g_0=0$ or $h_{ab}=h\d_{ab}$, they pair up and acquire non-zero energy. Being interested in the generic properties of the model, we can safely ignore symmetry-related zero-modes, which are massless at measure-zero regions in the parameter space.} $\iW(\bX^a):=(n_{\sss B}{-}n_{\sss F})=(2{-}0)=2$.

In the $\w_L \to0$ limit, the \Eqs{eZIn1}{eZIn2} become:
\begin{alignat}{7}
 \ddot x  &=0~,\quad&\To\quad   x(\t) &= v_{x \, 0}\,\t+ x_0~, \qquad&\quad
 \dot\c&=0~,\quad&\To\quad     \c(\t) &=\c_0~, \label{eZIn10}\\
 \ddot y &=0~,\quad&\To\quad    y(\t) &= v_{y \, 0}\,\t+ y_0~,\qquad&\quad
 \dot\h&=0~,\quad&\To\quad     \h(\t) &=\h_0~. \label{eZIn20}
\end{alignat}
Since \Eqs{eZIn10}{eZIn20} are {\em uncoupled\/}, the four bosonic, $( v_{x \, 0}, x_0, v_{y \, 0}, y_0)$, and two fermionic, $(\c_0,\h_0)$, integration constants remain independent. The number of {\em massless\/} on-shell degrees of freedom (zero-modes) then is: $n_{\sss B}=4$, and $n_{\sss F}=2$, leaving the Witten index at $(n_{\sss B}{-}n_{\sss F})=2$. While the Witten index remains unchanged in the $\w_L\to0$ limit, its separate contributions, $n_{\sss B}$ and $n_{\sss F}$, do change: This follows the original mode-migration wisdom\cite{rEWind}, and is presented graphically below:
\begin{equation}\setlength{\unitlength}{.8mm}
 \vC{\begin{picture}(110,30)(0,0)
  \put(0,0){\includegraphics[height=30\unitlength]{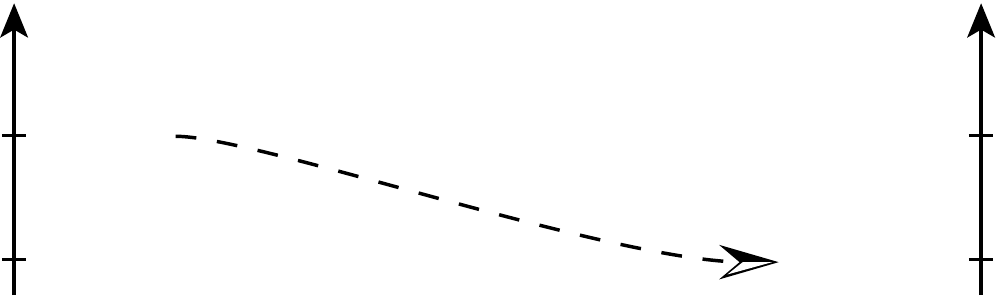}}
   \put(-26,3){$\iW=2$}
   \put(-6,27){$B$}
   \put(5,27){$F$}
   \put(-6,15){$2$}
   \put(6,15){$2$}
   \put(-6,3){$2$}
   \put(6,3){$0$}
   \put(50,12){$\w_L\propto\cB_0\to0$}
   \put(92,27){$B$}
   \put(103,27){$F$}
   \put(92,15){$0$}
   \put(104,15){$0$}
   \put(92,3){$4$}
   \put(104,3){$2$}
   \put(112,3){$\iW=2$}
 \end{picture}}
 \label{eMigA}
\end{equation}
Conversely, turning the magnetic field on, the degeneracy among the zero modes is partially ``lifted'', producing a supersymmetric Zeeman effect and leaving only two bosonic zero-modes.

\paragraph{The $\bJ^\rA$:} 
The Lagrangian\eq{eLF} produces the equations of motion $\dot\j^\rA=0$ and $F^\rA=0$, solved by $d_F{+}1$ fermionic constants, see \Eq{eN1F'}. This implies that the Witten index is $\iW(\bJ^\rA)=-3$ if $\d_0\neq0$, and $\iW(\bJ^\rA)=-2$ if we set $\d_0=0$ and drop the $\bJ^0$ superfield.

\subsubsection{$\bX^a$-$\bJ^\rA$ Mixing}
Using\eq{eBcD} and\eq{eFcD}, it is easy to see that the most general mixing and interactions between the fermionic and bosonic supermultiplets $\bX^a$ and $\bJ^A$ can be introduced via:
\begin{equation}
 \sL_\text{Int}~=~D~\big\{\, \cW(\bJ,\bX)\,\big\}\big|
  ~=~\j^A\,\cW\!,_{A}(\j,x)~-~i\c^a\,\cW\!,_a(\j,x)~,
 \label{eW(f,b)}
\end{equation}
where $\cW(\bJ,\bX)$ is a {\em\/fermionic\/} function of its arguments, and
$\cW\!,_{A}$ and $\cW\!,_a$ its left-derivatives by $\bJ^\rA$ and $\bX^a$, respectively. Herein, we focus on the simple, bilinear mixing terms\eq{eLBF}.

The combination $\sL_\text{B}+\sL_\text{L}+\sL_\text{F}+\sL_\text{B$\cdot$F}$, as given in\eqs{eLB}{eLBF}, produces the equations of motion, $F^a=-\w_0\,x^a$, the use of which produces:
\begin{equation}
 \begin{aligned}
 \sL_\text{B+F}\big|_{F^a}
 &= \inv2\,\d_{ab}\,(\,\dot x^a\,\dot x^b\,-\,{\w_0}^2\,x^a\,x^b\,)
    ~+~\inv2\,\w_L\,\ve_{ab}(\,x^a\,\dot x^b\,\,-\,i\,\c^a\,\c^b\,)\\*[1mm]
  &\mkern25mu+\frc{i}2\,\d_{ab}\,(\,\c^a\,\dot\c^b\,+\,\j^a\,\dot\j^b\,)
             ~+~i\,\w_0\,\d_{ab}\,\j^a\,\c^b~.
\end{aligned}
\label{eB+F}
\end{equation}
This describes a 2-dimensional, $N=1$-supersymmetric harmonic oscillator coupled to an external magnetic field. In particular, notice that the mixing parameter $\w_0$ introduced in\eq{eLBF} turns into the (radial) characteristic frequency of this oscillator.

Higher-dimensional generalizations, with $d>2$ superfields $\bX^a,\bJ^b$, will similarly describe supersymmetric, $d$-dimensional harmonic oscillators coupled to a higher-dimensional external magnetic field, $(\sF_0)_{ab}$ replacing the Larmor frequency coefficient, $\w_L\,\ve_{ab}$ in\eq{eLBF}.

\paragraph{The $\w_L,\w_0=0$ Case:}
This is the free-field limit where
\begin{equation}
  x^a = v_0^a\,\t+ x_0^a~,\qquad
 \c^a =\c_0^a ~,\quad \j^a\,=\, \j_0^a  ~,\qquad a=1,2~. \label{eSfHH0}
\end{equation}
Counting the bosonic and fermionic constants, the Witten index is, formally, $\iW=(4{-}4)=0$.

\paragraph{The $\w_L=0$, $\w_0\neq0$ Case:}
Now, the solutions take the form:
\begin{align}
  x^a &= x_+^a \cos(\w_0 \t)+ x_-^a \sin(\w_0 \t)~,\\
 \c^a &=\c_+^a \cos(\w_0 \t)+\c_-^a \sin(\w_0\t)~,\\
 \j^a &=\c_-^a \cos(\w_0 \t)-\c_+^a \sin(\w_0 \t)~, \label{eSHH}
\end{align}
and all constants of integration are associated with modes of nonzero frequency $\w_0$, so that $\iW=(0{-}0)=0$. 
 Thus the mode-migration diagram is
\begin{equation}\setlength{\unitlength}{.8mm}
 \vC{\begin{picture}(110,32)(0,0)
  \put(0,0){\includegraphics[height=30\unitlength]{SS11.pdf}}
   \put(-26,3){$\iW=0$}
   \put(-6,27){$B$}
   \put(5,27){$F$}
   \put(-6,15){$4$}
   \put(6,15){$4$}
   \put(-6,3){$0$}
   \put(6,3){$0$}
   \put(45,17){$\begin{aligned}
                \w_L&=0\\[-1mm]
                \w_0&\to0
               \end{aligned}$}
   \put(92,27){$B$}
   \put(103,27){$F$}
   \put(92,15){$0$}
   \put(104,15){$0$}
   \put(92,3){$4$}
   \put(104,3){$4$}
   \put(112,3){$\iW=0$}
 \end{picture}}
 \label{eMigB}
\end{equation}
and the difference between\eq{eMigB} and\eq{eMigA} owes to the introduction of
the $\j^a$ fermions.

\paragraph{The $\w_L,\w_0\neq0$ Case:}
The equations of motion are:
\begin{subequations}
 \label{e}
\begin{align}
 \ddot x^a~-~\w_L\,\ve^a{}_b\,\dot x^b ~+~ {\w_0}^2\,x^a~&=~0~,\label{eSHHx}\\
 \dot\c^a~-~\w_L\,\ve^a{}_b\,\c^b ~-~ \w_0\,\j^a ~&=~ 0 ~, \label{eSHHj1}\\
 \dot\j^a~+~ \w_0\, \c^a ~&=~ 0~. \label{eSHHj2}
\end{align}
\end{subequations}
With the help of\eq{eSHHj2}, the time derivative of\eq{eSHHj1} is:
\begin{align}
 \ddot \c^a &\, -~ \w_L\,\e^{a\, b}\d_{b\,c}\,\dot\c^c ~+~ {\w_0}^2\,\c^a ~=~ 0~,
\label{eSHBHrm}
\end{align}
which also follows as the supersymmetry variation of \Eq{eSHHx}. Since\eq{eSHBHrm} is identical in form to\eq{eSHHx}, so will be the on-shell solutions for $x^a$ and $\c^a$. It then suffices to discuss only the explicit form of the bosonic solution:
\begin{equation}
 x^a(\t)
   ~=~ A^a_+\,\cos(\w_+\t)+A^a_-\,\cos(\w_-\t)
      +\ve^a{}_b\,A^b_+\sin(\w_+\t)+\ve^a{}_b\,A^b_-\sin(\w_-\t)~,\label{eBosolut}
\end{equation}
where $\w_\pm=\frac12(\sqrt{\w_L^{~2}+4\w_0^{~2}}\pm\w_L)\geq0$.
This solution indicates that there are two constants of integration associated with the
frequency ${\w_+} $ and two with ${\w_-} $.  Thus, the mode-migration diagram becomes:
\begin{equation}\setlength{\unitlength}{.8mm}
 \vC{\begin{picture}(110,32)(0,0)
  \put(0,0){\includegraphics[height=30\unitlength]{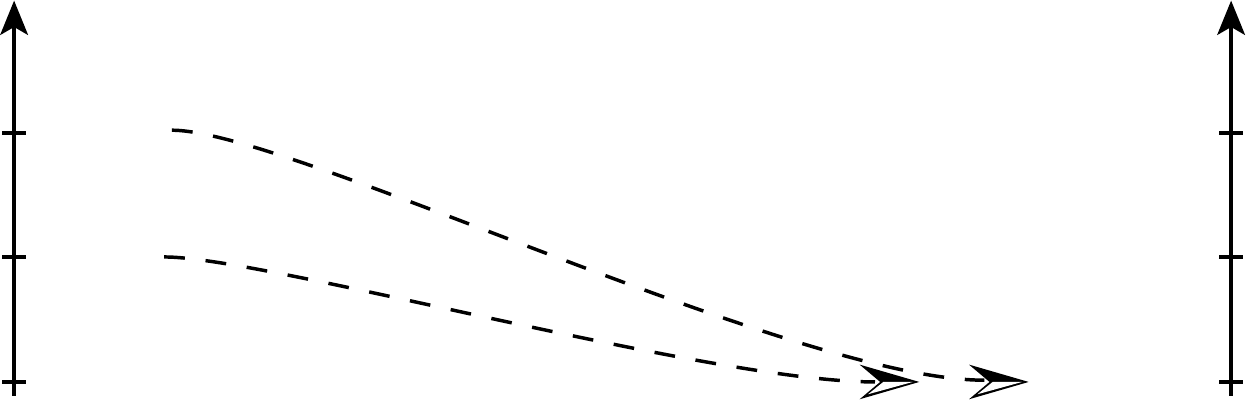}}
   \put(-26,1){$\iW=0$}
   \put(-6,27){$B$}
   \put(5,27){$F$}
   \put(-6,19){$2$}
   \put(6,19){$2$}
   \put(-6,10){$2$}
   \put(6,10){$2$}
   \put(-6,1){$0$}
   \put(6,1){$0$}
   \put(45,17){$\begin{aligned}
                \w_L&\to0\\[-1mm]
                \w_0&\to0
               \end{aligned}$}
   \put(85,27){$B$}
   \put(96,27){$F$}
   \put(85,19){$0$}
   \put(97,19){$0$}
   \put(85,10){$0$}
   \put(97,10){$0$}
   \put(85,1){$4$}
   \put(97,1){$4$}
   \put(103,1){$\iW=0$}
 \end{picture}}
 \label{eMigC}
\end{equation}
In the first, physical quadrant of the $(\w_0,\w_L)$-plane, the diagram\eq{eMigC} depicts a diagonal path, while\eq{eMigB} follows the $\w_0$-axis.

On the other hand, along the $\w_L$-axis we then have
\begin{equation}\setlength{\unitlength}{.8mm}
 \vC{\begin{picture}(110,32)(0,0)
  \put(0,0){\includegraphics[height=30\unitlength]{SS11.pdf}}
   \put(-26,3){$\iW=0$}
   \put(-6,27){$B$}
   \put(5,27){$F$}
   \put(-6,15){$2$}
   \put(6,15){$2$}
   \put(-6,3){$2$}
   \put(6,3){{$2$} $(\j^a)$}
   \put(45,17){$\begin{aligned}
                \w_L&\to0\\[-1mm]
                \w_0&=0
               \end{aligned}$}
   \put(92,27){$B$}
   \put(103,27){$F$}
   \put(92,15){$0$}
   \put(104,15){$0$}
   \put(92,3){$4$}
   \put(104,3){$4$}
   \put(112,3){$\iW=0$}
 \end{picture}}
 \label{eMigD}
\end{equation}
Note, however, that the $\bJ^a$-modes are {\em\/decoupled\/} from the $\bX^a$-modes along the $\w_0=0$ edge.

The mode-migration diagrams\eq{eMigB},\eq{eMigC} and\eq{eMigD} show that $\w_L$ partially lifts the degeneracy of the modes, in a Zeeman-like response to an external magnetic field.

\subsection{Obstruction to Supersymmetry Breaking}
 \label{sSSBO}
The combination of\eq{eMigB}--(\ref{eMigC})--(\ref{eMigD}) then covers the behavior in the 1st quadrant, the physical region, of the $(\w_L,\w_0)$-plane. The Witten index of the system\eq{eB+F}, $\iW=0$, remains constant throughout the physical region of the $(\w_0,\w_L)$ parameter space of this system.

However, in the $\w_0=0=g_0$ subregion of the parameter space, the system\eq{eB+F} decouples into two separate sub-systems: the $\bX^a$-system\eqs{eLB}{eLL} with $\iW(1)=+2$, and the $\bJ^a$-system\eq{eLF} with $\iW(2)=-2$, having set $\d_0=0$ and having dropped $\bJ^0$. As the $g\neq0$ interactions cannot induce any mode-pairing while $\w_0=0$, this obstruction to supersymmetry breaking---{\em\/finer\/} than the overall Witten index\cite{rEWind}---is limited only to this, $\w_0=0$, ``unmixing'' edge of the parameter space.

Therefore, any and all supersymmetry breaking effects in the ``bulk'' of the whole parameter space must: ({\bf1})~vanish in the $\w_0\to0$ limit, and if necessary, ({\bf2})~be discontinuous in this limit.

\subsection{Supersymmetry Breaking}
As far as we know, there are no general guarantees that a supersymmetric model with a non-vanishing Witten index can be embedded into another supersymmetric model with a vanishing Witten index. A result to this effect would seem to be of interest, especially because a nonzero Witten index is understood to obstruct supersymmetry breaking\cite{rEWind}.

The obvious embedding of\eqs{eLB}{eLL} into\eq{eB+F} is precisely an example of such an embedding.

So, whereas the standard argument\cite{rEWind} prohibits supersymmetry breaking in a model that limits to\eqs{eLB}{eLL}, the same argument permits supersymmetry breaking in its augmentations that limit to\eq{eB+F}---except at the $\w_0=0$ edge, where the augmentation ``unmixes''.

Since many physics models involve constrained target subspaces, we now return to the full Lagrangian\eqs{eLB}{eLC}.

\subsubsection{The $\d_0=0$ Case:}
 Upon eliminating the auxiliary fields $F^a$, the Lagrangian becomes
\begin{align}
 \Tw{\sL}_\text{B+F}\big|_{F^a}
&= \inv2\,\d_{ab}\,(\,\dot x^a\,\dot x^b\,-\,{\w_0}^2\,x^a\,x^b\,)
    ~+~\inv2\,\w_L\,\ve_{ab}(\,x^a\,\dot x^b\,\,-\,i\,\c^a\,\c^b\,)\nn\\*[1mm]
 &\mkern25mu+~\inv2\,i\,\d_{ab}\,(\,\c^a\,\dot\c^b\,+\,\j^a\,\dot\j^b\,)
           ~+~i\,\w_0\,\d_{ab}\,\j^a\,\c^b\nn\\*[1mm]
 &\mkern25mu+~\inv2\,g_0\, F^0\, \big(\, x^a\,h_{ab}\,x^b - R^2\,\big)
           ~+~ i\,g_0\,\j^0\,( x^a\,h_{ab}\,\c^b)~,\label{eB+FCon}
\end{align}
and the equations of motion now become:
\begin{align}
 \ddot x^a ~-~ \w_L\,\ve^a{}_b\,\dot x^b
           ~+~ \big(\, {\w_0}^2\d^a{}_b - g_0\, F^0\,h^a{}_b\,\big)\,x^b
           ~-~ i\,g_0\,\j^0\,h^a{}_b\c^b &=~0~,
 \label{eConx}\\*[2mm]
 \dot \c^a ~-~ \w_L\,\ve^a{}_b\,\c^b  ~-~ \w_0\,\j^a
           ~-~ g_0\,\j^0\,h^a{}_b\,x^b &=~0~,
 \label{eConc}   \\[2mm]
 \dot \j^a ~+~ \w_0\,\c^a &=~0~, \label{eConj} \\[2mm]
 x^a\,h_{ab}\,x^b ~-~ R^2 &=~ 0 ~, \label{eConF0}\\[2mm]
 x^a\,h_{ab}\,\c^b &=~ 0~, \label{eConj0}
\end{align}
where $\ve^a{}_b:=\d^{ac}\,\ve_{cb}$ and $h^a{}_b:=\d^{ac}\,h_{cb}$.

\paragraph{Algebraic Constraints:}
\Eq{eConF0} implies that $\|x\|_h^2=R^2$, constraining the two bosons to this quadratic curve $\sC\subset\IR^2$.
 This curve, $\sC$, is an ellipse if $h_{ab}$ is positive definite, a hyperbola if $h_{ab}$ has eigenvalues of both signs, and a straight line if one of the eigenvalues vanishes and the other is positive. Otherwise, $\sC=\varnothing$.
 \Eq{eConj0} implies that $\c^b$ are $h_{ab}$-orthogonal to $x^a$, \ie, tangential to $\sC$ at each of its points. Thus, the $\c^b$ span the fibers of the tangent bundle, $T_\sC$.

\paragraph{Dynamics:}
One of the 2 equations\eq{eConx} may be used to express $F^0$ in terms of the other fields; the remaining equation governs the dynamics of the $x^a$, constrained to $\sC$.
 One of the 2 equations\eq{eConc} may be used to express $\j^0$ in terms of the other fields, and the remaining equation governs the dynamics of the $\c^a$, constrained to the fibers of $T_\sC$. Finally, Eqs.\eq{eConj} relate a combination of the $\j^a$'s to a corresponding $\c^a|_{T_\sC}$ as a matching pair to\eq{eConc}, while the other combination of the $\j^a$'s is restricted from varying away from $T^*_\sC$.

\paragraph{Example:}
By selecting $h_{22}=1$ to be the only non-zero element of $h_{ab}$, we obtain:
\begin{gather}
 x^2=\pm R~,\quad \c^2=0~,\qquad
 F^0=\frc{\pm1}{g_0R}(\w_L\,\dot{x}^1\pm{\w_0}^2\,R)~,\qquad
 \j^0=\frc{\pm1}{g_0R}(\w_L\,\c^1-\w_0\,\j^2_0)~,\\[1mm]
 \ddot{x}^1+{\w_0}^2\,x^1=0~,\qquad
 \dot\c^1-\w_0\,\j^1=0=\dot\j^1+\w_0\c^1~,\qquad
 \j^2=\text{\it const.}\qquad
\end{gather}
Here $\sC$ consists of two copies of the $x^1\id x$-axis, positioned at $x^2\id y=\pm R$. The fermions $\c^1\id\c$ and $\j^1$ span $T_\sC$ and $T^*_\sC$, respectively\Ft{In fact the supersymmetry transformations also imply this: $Q:x^a\to\c^a$, whereas $Q:\j^a\to F^a=-\w_0 x^a$.}, $\c^2=0$, $\j^2=\text{\it const.}$, and $F^0,\j^0$ are functions of other fields.

\paragraph{Supersymmetry Breaking:}
The bosonic potential is obtained from the negative of\eq{eB+FCon}, by setting all fermions and all $\t$-derivatives to zero, and enforcing the constraint\eq{eConF0}:
\begin{equation}
 \sV(x|_\sC) = \inv2\,{\w_0}^2(x^a\,\d_{ab}\,x^b)\big|_{\|x\|_h^2=R^2}~. \label{ePoth}
\end{equation}
Now, if we choose $h_{ab}=h\,\d_{ab}$, the constraint\eq{eConF0} implies that
\begin{equation}
 \sV(x|_\sC)=\inv2\,{\w_0}^2\,R^2/h~>0~,\quad\text{since}\quad
  \sign(h)\isBy{(\ref{eConF0})}\sign(R^2)~. \label{eV>0}
\end{equation}
This precludes the total energy of the system from vanishing, and so also the existence of supersymmetric ground states: supersymmetry is spontaneously broken. The same holds for other choices of $h_{ab}$, regardless of its (in)definiteness, as long as $R\neq0$. In fact, even in the analytic continuation to $R^2<0$ {\em\/\'a~la\/} Witten\cite{rPhases}, the result\eq{eV>0} remains true and $R\neq0$ breaks supersymmetry.

Recall now that the $\w_0=0$ edge of the physical parameter space harbors the formal obstruction to supersymmetry breaking. Indeed $\lim_{\w_0\to0}\sV(x|_\sC)=0$, and supersymmetry breaking is turned off in the $\w_0\to0$ limit. The same is true in the $R\to0$ limit.

In turn, neither $\sV(x)$ nor $\sV(x|_\sC)$ depend on $\w_L$: the coupling to the external magnetic field has no effect on supersymmetry breaking, and the Zeeman effect is supersymmetric.

\subsubsection{The $\d_0\neq0$ Case:}
The effect of $\d_0\neq0$ is that $\j^0$ also becomes a dynamical fermion, while $F^0$ still has a purely algebraic equation of motion: $F^0=-\inv2(g_0/\d_0)\big(\|x\|_h^2-R^2\big)$. Upon substituting this back into the Lagrangian, the potential becomes
\begin{equation}
 \sV(x)=\inv2{\w_0}^2(x^a\,\d_{ab}\,x^b)
      ~+~\inv8\,\Big(\frac{g_0}{\d_0}\Big)^2\Big((x^a\,h_{ab}\,x^b)-R^2\Big)^2
 \label{ePot}
\end{equation}
When $h_{ab}=\d_{ab}$, the extrema of this include the origin, $x^a=0$ and the circle of radius $\|x\|=\sqrt{R^2-2(\d_0\,\w_0/g_0)^2}$ when $R\geq\sqrt2\,\d_0\,\w_0/g_0$, breaking the $SO(2)$ symmetry\Ft{We forego gauging this as well as other symmetries that emerge in special regions of the parameter space.}. Throughout, however, $\sV(x)$ remains positive---{\em\/signaling supersymmetry breaking\/}---except when $\w_0\to0$. The graph
\begin{equation}
 \vC{\begin{picture}(140,50)(0,0)
  \put(10,0){\includegraphics[height=50\unitlength]{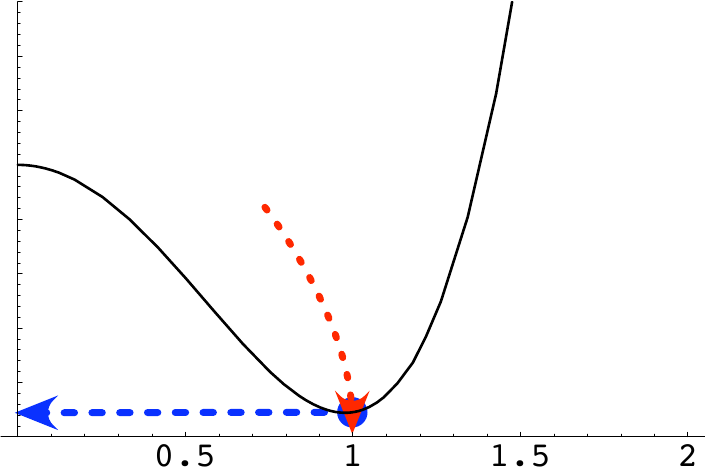}}
  \put(13,45){$\sV(x)$}
  \put(73,6){$\|x\|/R$}
  \put(13,8){\footnotesize$R\leq \sqrt2\,\d_0\,\w_0/g_0$}
  \put(42,24){\footnotesize$\w_0\to0$}
  \put(90,40){\parbox[t]{49\unitlength}%
              {Supersymmetry is restored in the $\w_0\to0$ limit.}}
  \put(90,24){\parbox[t]{42\unitlength}%
              {$SO(2)$ rotational symmetry is restored when $R\leq \sqrt2\,\d_0\,\w_0/g_0$.}}
 \end{picture}}
 \label{ePotPic}
\end{equation}
shows effect of the variation of $\w_0$ and $R$ on the potential $\sV(x)$; see\eq{ePot}. This r\^ole of $\w_0\to0$ as the supersymmetry restoration limit is perfectly in agreement with the above analysis of the obstruction to supersymmetry breaking in section~\ref{sSSBO}.

\section{$N>1$ Isoscalar and Isospinor Supermultiplets}
 \label{sN>1}
With the foregoing analysis of the $N=1$ case, we now turn to the much more interesting, {\em\/arbitrary\/} $N>1$ generalization.

The transformation rules of the {\em Isoscalar\/} supermultiplet\cite{r6-2} may be written as:
\begin{equation}
  Q_\rI\,x_i    =(\,L_\rI\,)_i\,^\hj\,\c_\hj~,\qquad
  Q_\rI\,\c_\hi =i\,(\,R_\rI\,)_\hi\,^j\,\dot x_j~.
 \label{eISc}
\end{equation}
which straightforwardly generalizes\eq{eISN1}.
 Although the number of bosonic and fermionic component fields is the same, $2^{N-1}$, we find it useful to distinguish between the indices, $i$ {\em vs\/}.\ $\hi$, that count them. Finally, we note that such a supermultiplet has the {\em topology\/}\cite{r6-1} of the $N$-cube, $[0,1]^N$. Smaller, {\em\/quotient\/} supermultiplets may be obtained using certain {\em projections\/}, as classified in Ref.\cite{r6-3}. Including also these quotient supermultiplets provides an extension of our present analysis, but is beyond our present scope.

For consistency with\eq{eSuSy}, the $\IL_\rI$ and $\IR_\rI$ matrices in\eq{eISc} must satisfy
\begin{subequations}
 \label{eRL}
 \begin{alignat}{3}
 (\,L_\rI\,)_i{}^\hj\>(\,R_\rJ\,)_\hj{}^k + (\,L_\rJ\,)_i{}^\hj\>(\,R_\rI\,)_\hj{}^k
  &= 2\,\d_{\rI\rJ}\,\d_i{}^k~,\qquad&\text{\ie},\qquad
 \IL_\rI\,\IR_\rJ+\IL_\rJ\,\IR_\rI&=2\,\d_{\rI\rJ}\,\Ione~;\\*[1mm]
 (\,R_\rJ\,)_\hi{}^j\>(\,L_\rI\,)_j{}^\hk + (\,R_\rI\,)_\hi{}^j\>(\,L_\rJ\,)_j{}^\hk
  &= 2\,\d_{\rI\rJ}\,\d_\hi{}^\hk~,\qquad&\text{\ie},\qquad
 \IR_\rI\,\IL_\rJ+\IR_\rJ\,\IL_\rI&=2\,\d_{\rI\rJ}\,\Ione~.
\end{alignat}
\end{subequations}
The $J=I$ cases then imply that
\begin{equation}
 \left.\begin{aligned}
  (\,L_\rI\,)_i{}^\hj\,(\,R_\rI\,)_\hj{}^k&=\d_i{}^k\\
  (\,R_\rI\,)_\hi{}^j\,(\,L_\rI\,)_j{}^\hk&=\d_\hi{}^\hk
  \end{aligned}\quad\right\}
  \quad\text{\ie,}\quad \IR_\rI=\IL_\rI^{-1}~,\qquad I=1,\cdots,N~.
 \label{eR=L-1}
\end{equation}

Generalizing similarly\eq{eN1F'}, we introduce {\itshape Isospinor\/} supermultiplets, $\bJ^A=(\j_\hi^A,F_i^A)$:
\begin{equation}
  Q_\rI\,\j_\hi =i\,(\,R_\rI\,)_\hi\,^j\,F_j~,\qquad
  Q_\rI\,F_i    =(\,L_\rI\,)_i\,^\hj\,\dot\j_\hj~.
 \label{eISp}
\end{equation}

\subsection{The Lagrangian}
Supersymmetry of the standard kinetic terms, generalizing\eq{eLB}, implies the relation\Ft{The difference with respect to the original relation\cite{rGR1,rGR2} owes to an overall sign-convention. Our present convention keeps the forms of\eq{eISc},\eq{eISp} in the $N\to1$ limit, and so also that of corresponding Lagrangians.}:
\begin{equation}
 \IR_\rI = \IL_\rI^{~T}~, \quad\text{\ie}\quad
 (\,R_\rI\,)_\hj{}^k\,\d_{ik} = (\,L_\rI\,)_i{}^\hk\,\d_{\hj\hk}~.
 \label{eR=LT}
\end{equation}
The conditions in\eq{eRL},\eq{eR=L-1} and\eq{eR=LT} define the ${\cal{GR}}(\rd, \,{\cal N})$ `Garden' algebras introduced in\cite{rGR1,rGR2}.

For $N>1$, these are {\em not\/} the familiar, Salam-Strathdee superfields of $N$-extended supersymmetry, although Theorem~7.6 in Ref.\cite{r6-1} relates them\Ft{Ref.\cite{r6-2} translates the standard kinetic terms into superfield notation. We see no obstruction to doing so also for our complete Lagrangian\eq{eSZEM-N}.}.
 In particular, in the Isoscalar supermultiplet $\bX^a$, all bosons $x^a_i$ have the same engineering dimension, $\inv2$ less than the fermions, $\c^a_\hi$. Consequently, given a suitable Lagrangian, all bosons and all fermions are physical, propagating component fields: There are neither auxiliary nor gauge degrees of freedom in $\bX^a$; this is also true of its quotient supermultiplets, mentioned above.

Generalizing\eq{eLL}, we seek an $N$-supersymmetric term of the general form:
\begin{gather}
 \sL_\text{L} = \inv2\,\big\{\,\w^{ij}_{ab}\,x^a_i\,\dot x^b_j
                            -i\,\ha\w^{\hi\hj}_{ab}\,\c^a_\hi\,\c^b_\hj\,\big\}~,
 \label{eAM}
\InT{with}
 \w^{ij}_{ab}=-\w^{ji}_{ba}~,\quad\text{and}\quad
  \ha\w^{\hi\hj}_{ab}=-\ha\w^{\hj\hi}_{ba}~,
 \label{eASym}
\end{gather}
and require that it be invariant with respect to the supersymetry transformation\eq{eISc}.
This is the case if and only if:
\begin{equation}
 (L_\rI)_i{}^\hk\,\w^{i\ell}_{ab}~=~\ha\w^{\hk\hj}_{ab}\,(R_\rI)_\hj{}^\ell~,
 \qquad \text{for all}~ a,b,~\ell,\hk,~I~.
 \label{eCondW}
\end{equation}

The condition\eq{eCondW} simplifies in the special case, when there is an even number, $d=2p$, supermultiplets $\bX^a$: Then, we divide the supermultiplets $\bX^a\to\bX^{\a\,\ah}$ into pairs, so that $\a=1,2$ and $\ah=\lfloor\frac{a+1}2\rfloor=1,\cdots,p$. The properties\eq{eASym} are then satisfied by choosing:
\begin{equation}
 \w^{ij}_{ab}=\ve_{\a\b}\,\d^{ij}\,\w_{(\ah\bh)}~,\quad\text{and}\quad
  \ha\w^{\hi\hj}_{ab}=\ve_{\a\b}\,\d^{\hi\hj}\,\ha\w_{(\ah\bh)}~.
 \label{eCh1}
\end{equation}
Setting then, in addition, $\w_{(\ah\bh)}=\ha\w_{(\ah\bh)}$, with\eq{eCh1} ensures the supersymmetry of the Lagrangian\eq{eAM}.

Together with the standard kinetic terms, this produces
\begin{align}
 \sL_\text{B}+\sL_\text{L}
 &=\inv2\,\d_{\ah\bh}\,\d_{\a\b}\,\big\{\,
           \d^{ij}\,\dot x^{\a\,\ah}_i\,\dot x^{\b\,\bh}_j
          +i\,\d^{\hi\hj}\,\c^{\a\,\ah}_\hi\,\dot\c^{\b\,\bh}_\hj\,\big\}\nn\\
 &\quad+\inv2\,\w_{(\ah\bh)}\big\{\,
       \d^{ij}\big(x^{1\,\ah}_i\,\dot x^{2\,\bh}_j
                                 -\dot x^{1\,\ah}_i\,x^{2\,\bh}_j\big)
          -2i\,\d^{\hi\hj}\,
                \c^{1\,\ah}_\hi\,\c^{2\,\bh}_\hj\,\big\}~, \label{eBN}
\end{align}
generalizing\eqs{eLB}{eLL}.

The structure of the matrix of Larmor frequencies\eq{eCh1} implies that the supermultiplets $\bX^{\a\,\ah}=(x_i^{\a\ah}|\c_i^{\a\ah})$ are organized into $p$ pairs, $(\bX^{1\,\ah},\bX^{2\,\ah})$. Each $\bX^{1\,\ah}$ then mixes with each $\bX^{2\,\bh}$ by an amount controlled by the Larmor frequency $\w_{(\ah\bh)}$. In turn, the ``trace'' angular momentum,
\begin{equation}
 \rL^{(\ah\bh)}:=\d^{ij}\,
 \big(x^{1\,\ah}_i\,\dot x^{2\,\bh}_j-\dot x^{1\,\ah}_i\,x^{2\,\bh}_j\big)~,
\end{equation}
couples, with the strength of $\w_{(\ah\bh)}$, to the external magnetic field through the $(x^{1\,\ah}_i,x^{2\,\bh}_i)$-planes.

\subsection{The $N>1$ Model}
We finally arrive at:
\begin{align}
 \sL_\text{SZEM}
  &=\sL_\text{B}+\sL_\text{L}+\sL_\text{F}+\sL_\text{B$\cdot$F}+\sL_\text{C}~,
  \label{eSZEM-N}
\intertext{where $\sL_\text{B}+\sL_\text{L}$ are given in\eq{eBN}, and}
 \sL_\text{F}~
  &=~\inv2\,\d_0\,\big(\,\d^{ij}\,F_i^0F_j^0
                              ~+~i\,\d^{\hi\hj}\,\j_\hi^0\,\dot\j_\hj^0\,\big)
    +\inv2\,\d_{ab}\,\big(\,\d^{ij}\,F_i^aF_j^b
                              ~+~i\,\d^{\hi\hj}\,\j_\hi^a\,\dot\j_\hj^b\,\big)~,
 \label{eFN}\\[2mm]
 \sL_\text{B$\cdot$F}
 &=~\w_0\,\d_{ab} \big(\,\d^{ij}\,F_i^a\, x_j^b
                        ~+~ i\,\d^{\hi\hj}\j_\hi^a\,\c_\hj^b\,\,\big)~,
 \label{eB.FN}\\[2mm]
\sL_\text{C}
 &=g_0\,\big\{\,\inv2\,F_i^0\big(\D^{i\,j\,k}_{~a\,b}\,x_j^a\,x_k^b-\D^i\,R^2\big)
           ~+~i\,\j_\hi^0\,(\Dh^{\hi\,j\,\hk}_{~a\,b}\,x_j^a\,\c_\hk^b)\,\big\}~.
\label{eConN}
\end{align}

\paragraph{Supersymmetry:}
Owing to the fact that each auxiliary field $F_i^A$ transforms into a total $\t$-derivative, the linear term $\inv2 g_0F^0_i\D^iR^2$ is supersymmetric all by itself for any dimensionless $2^{N-1}$-vector, $\D^i$.

By construction of\eq{eConN},
\begin{subequations}\label{eConD}
 \begin{gather}
 \D^{i\,j\,k}_{~a\,b} = \D^{i\,k\,j}_{~b\,a}~. \label{eConD0}
\intertext{The arbitrary $N$-extended supersymmetry of\eq{eConN} is then ensured if the arrays of dimensionless constants $\D^{i\,j\,k}_{~a\,b}$ and $\Dh^{\hi\,j\,\hk}_{~a\,b}$ satisfy:}
 (R_\rI)_\hl{}^i\,\Dh^{\hl\,j\,\hk}_{~a\,b}
 =\D^{i\,j\,m}_{~a\,b}\,(L_\rI)_m{}^\hk~,\qquad
 \Dh^{\hi\,j\,\hl}_{~a\,b}\,(R_\rI)_\hl{}^k
 =(L_\rI)_\ell{}^\hi\,\D^{\ell\,j\,k}_{~a\,b}~, \label{eConD1}\\
 \Dh^{\hi\,\ell\,\hk}_{~a\,b}\,(L_\rI)_\ell{}^\hj
 =\Dh^{\hi\,\ell\,\hj}_{~b\,a}\,(L_\rI)_\ell{}^\hk~. \label{eConD2}
\end{gather}
\end{subequations}
Using\eq{eR=LT}, the equations\eq{eConD1} produce
\begin{equation}
   \Dh^{\hi\,j\,\hk}_{~a\,b}
 = \inv{N}\sum_{I=1}^N (L_\rI)_\ell{}^\hi\,\D^{\ell\,j\,m}_{~a\,b}\,(L_\rI)_m{}^\hk~,
 \label{eDefD}
\end{equation}
which may be used as a {\em\/definition\/} of $\Dh^{\hi\,j\,\hk}_{~a\,b}$ in terms of $\D^{i\,j\,k}_{~a\,b}$. In fact, the conditions\eqs{eConD1}{eConD2} are all satisfied if the $N$ contributions in the defining sum\eq{eDefD} are all identical.

As the system of constraints\eq{eConD} may seem over-constraining, we exhibit the simplest non-trivial Ansatz to solve the conditions\eq{eConD} for $N=2$, where we choose
\begin{equation}
 \IL_1=\bsm 0&1\\[1pt] 1&0 \esm=\IR_1~,\qquad
 \IL_2=\bsm 1&0\\[1pt] 0&-1 \esm=\IR_2~, \label{ChRL}
\end{equation}
restrict $a,b=1,2$, and find
 \begin{align}
 (\D^{ijk}_{\>11})
 &=\big(\,\bsm A&~B\\[1pt]B&-A\esm\,,\,\bsm -B&A\\[1pt]~A&B\esm\,\big)~,
 \qquad\To\qquad
 (\Dh^{\hi j\hk}_{\>11})
  =\big(\,\bsm A&-B\\[1pt]B&~A\esm\,,\,\bsm ~B&A\\[1pt]-A&B\esm\,\big)~,\label{Ds}\\[2mm]
 &=\d^{j1}(A\,\d^{ik}+B\,\ve^{ik}) + \d^{j2}(B\,\d^{ik}-A\,\ve^{ik})~,
 \label{eGenDelta}
\end{align}
where $A,B$ are two arbitrary constants; the index $i$ labels the two blocks, in which $j,k$ are the row- and column-indices, respectively.
 $(\D^{ijk}_{\>12})$ and $(\Dh^{\hi j\hk}_{\>12})$ are of the same form, depending on another two arbitrary constants, and so are $(\D^{ijk}_{\>22})$ and $(\Dh^{\hi j\hk}_{\>22})$.

Note now that the $N=1$ constraint Lagrangian\eq{eLC} depended on 3 constants, $h_{ab}$, which is, for the $N=2$ constraint system\eq{eConN}, generalized to the 6 constants in $\D^{ijk}_{\>ab}$. These numbers precisely fit the generic formula one would expect, $N{\cdot}{d+1\choose2}$, where $N$ stems from the $N$-extendedness of supersymmetry, and ${d+1\choose2}$ is the number of parameters in a quadratic\Ft{For degree-$q$ polynomials, 
 $(\D^{ij_1\cdots\,j_q}_{\,a_1\cdots\,b_q})$ subject to a generalization of\eq{eConD} and\eq{eGenDelta2} would depend on $N{\cdot}{d+q-1\choose q}$ parameters; effectively, one degree-$q$ polynomial in $d$ coordinates for each of $N$ supersymmetries.} polynomial in $d$ coordinates.
We thus expect no obstruction to finding solutions to the conditions\eq{eConD} for arbitrary $N$ and $d$.

Looking at the solution above and using the representation theory of the ${\cal{GR}}(\rd, \,{\cal N})$ algebras, it seems practical to expect that there exist constant coefficients $\D_{a \, b \,\ell}$ such that
\begin{align}
 (\D^{ijk}_{\>a \, b}) ~&=~
 \sum_r \, 
   \Big(\D_{a\,b\,\ell}\,(f^{\rI_1\cdots\rI_{2r}})^{\ell\,j}\,
                         (f_{\rI_1\cdots\rI_{2r}})^{i\,k}
   +\D_{b\,a\,\ell}\,(f^{\rI_1\cdots\rI_{2r}})^{\ell\,k}\,
                         (f_{\rI_1\cdots\rI_{2r}})^{i\,j}\Big)~,
  \label{eGenDelta2}
\end{align}
with the Clebsch-Gordan-like coefficients $(f_{\rI_1\cdots\rI_{2\ell}})^{i\,k}$ defined in Ref.\cite{rGR2}, is the natural and covariant generalization of the $N=2$ specific result\eqs{Ds}{eGenDelta}.
Though additional computations remain to construct explicit such models for even higher values of $N$ and for which the conditions in \eq{eConD} are satisfied, this method does seem to have the potential to open a new study on whether it is possible to construct a model with  $N\to N'$-extended supersymmetry breaking, with $N>N'\neq0$.

\paragraph{The $\d_0=0$ Case:}
As in the $N=1$ case, the component fields of the supermultiplet $(\j^0_\hi|F^0_i)$ act as Lagrange multipliers. The auxiliary fields $F^0_i$ enforce the $N$ bosonic constraints
\begin{equation}
 \D^{i\,j\,k}_{~a\,b}\,x_j^a\,x_k^b=\D^i\,R^2~, \label{eCBBN}
\end{equation}
whereas the fermions, $\j^0_\hi$ impose $N$ $\D$-orthogonality constraints
\begin{equation}
 \Dh^{\hi\,j\,\hk}_{~a\,b}\,x_j^a\,\c_\hk^b = 0~. \label{eCBFN}
\end{equation}
A detailed analysis of the geometry of the so-defined target space is beyond our present scope, but it should be clear that the system\eqs{eCBBN}{eCBFN}, parametrized by the $N{\cdot}{d+1\choose2}$ parameters in $\D^{i\,j\,k}_{~a\,b}$ and the $2^{N-1}$ components of $\D^i$, offers considerable choices.

\paragraph{The $\d_0\neq0$ Case:}
The equations of motion for the fields $F^0_i$ now become
\begin{equation}
 F^0_i = -\frac{g_0}{2\d_0}\,\d_{ij}\,
           \big(\D^{j\,k\,\ell}_{~a\,b}\,x_k^a\,x_\ell^b-\D^j\,R^2\big)~,
\end{equation}
which, when substituted back into the Lagrangian produces the potential
\begin{equation}
 \sV(x)=\inv2{\w_0}^2(x_i^a\,\d^{ij}\d_{ab}\,x_j^b)
      ~+~\inv8\,\Big(\frac{g_0}{\d_0}\Big)^2
          \Big\|\big(\D^{i\,j\,k}_{~a\,b}\,x_j^a\,x_k^b-\D^i\,R^2\big)\Big\|^2~,
 \label{ePotN}
\end{equation}
providing the $N>1$ and $d$-dimensional generalization of\eq{ePot}.

\section{Conclusions}
 We have shown that it is possible to construct a new class of supersymmetical quantum mechanical models that admit an {\em\/arbitrary number of supercharges\/} and interactions with static, background magnetic fields (\ie, fluxes).  In the case of $N = 1$, we have seen how the mode migration of the model is in accord with the Witten index arguments\cite{rEWind}, but have also found that in a certain limit there is an obstruction to supersymmetry
 breaking that is finer than the one based on the (overall) index alone.
 For the case of $N=2$, we have also introduced explicit interactions, including some of a form very similar to Landau-Ginzburg models involving toric geometry\cite{rPhases}.
 We also provided a parameter-counting argument to indicate that, in fact, there is no obstruction to constructing the model\eq{eSZEM-N} for any $N\in\IN$.
 Our detailed analysis of the $N = 1$ model, embedded within the $N = 2$ one, implies that, for all $N$, the bulk of the parameter space describes a phase where supersymmetry is  spontaneously broken, but is restored in the boundary.
 
 The current study does not exhaust the class of models with interactions and arbitrary numbers of supersymmetries that may be constructed as generalizations of these techniques.  In particular, in future efforts it seems indicated that the special case of $N = 32$ may be of special interest, at which point the classification of Ref.\cite{r6-3} will help with the combinatorial complexity.
 
 The two pressing questions are whether such a model can be constructed that might allow a new method for the study of $M$-Theory, and whether  
the constrained target space may be tailored into a Ho\v{r}ava-Witten spacetime with boundary brane-Worlds\cite{rPHEW1,rPHEW2}, complete with a Randall-Sundrum geometry\cite{rRS1,rRS2}.

\bigskip
\begin{flushright}\sl
You cannot depend on your eyes when your imagination is out of focus.\\*[-1mm]
 --~Mark Twain
\end{flushright}

\bigskip\paragraph{Acknowledgments:}
 The research of S.J.G.\ is supported in part by the National Science Foundation Grant PHY-0354401, the endowment of the John S.~Toll Professorship and the CSPT.
 T.H.\ is indebted to the support of the Department of Energy through the grant DE-FG02-94ER-40854.

\bibliographystyle{elsart-numX}
\bibliography{Refs}

\end{document}